# Superconducting Cosmic String That Connected A Charged Black Hole and Considered as Hair of Black Hole

**Ali Riza AKÇAY**
Barbaros Mah. Veysipasa Sok. 100.
Yil Sitesi H-Blok No: 17/5, 34662
Uskudar, Istanbul, Turkey
Tel: + 90 542 321 66 89
E-mail: akcayar@e-kolay.net

**Abstract:** This paper describes that the superconducting cosmic strings can be connected to a charged black hole and can be considered as the hair of black hole. What the no-hair theorems show is that a large amount of information is lost when a body collapses to form a black hole. In addition, the no-hair theorem has not been proved for the Yang-Mills field. This paper proves and claims that the superconducting cosmic strings can be connected to a charged black hole when the current inside these strings and black holes approaches the critical value ($J \to J_c = emc^2/\hbar$). Because, this state is the final state of the gravitational collapse, and the event horizon would be destroyed in this state. Therefore, these strings should be considered as hair of the charged black holes and may be titled as BHCS (black hole connected strings). This means that at least the charged black holes have the hair. Thus, the no-hair theorem is not applicable for the charged black holes in the state of the critical current ($J \sim J_c$).

## 1. BACKGROUND

John Wheeler, who invented most of the terminology concerning black holes, remarked in the 1960s that 'black holes have no hair' [1]. The no-hair theorem proved by the combined work of Hawking, Israel, Carter and Robinson shows that the only stationary black holes in the absence of matter fields are the Kerr solutions. The no-hair theorem was extended by Robinson to the case where there was an electromagnetic field. This added a third parameter Q, the electric charge. The no-hair theorem has not been proved for the Yang-Mills field [2].

H. Reissner in 1916, and independently G. Nordström in 1918, discovered an exact solution to Einstein's equations for the gravitational field caused by an electrically charged mass. This solution is a generalized version of Schwarzschild's solution, with one other parameter: the electric charge. It describes space-time outside the event horizon of an electrically charged black hole. There is a limit to the amount of electric charge a black hole may have. Above a critical limit ($J > J_c = emc^2/\hbar$) the event horizon would be destroyed by the colossal force of electrostatic repulsion. The maximum electric force is proportional to the mass of black hole. Nevertheless, a black hole is just as likely to be positively charged as negatively charged [1].

The idea that strings could become superconducting was first suggested in a pioneering paper by Witten (1985a). Later it was realized that the role of the superconducting condensate could be played not only by a scalar field, but also by a vector field whose flux is trapped inside a non-abelian string (Preskill, 1985; Everett, 1988). If the vector field is charged, the gauge invariance is again spontaneously broken inside the string. Witten also proposed another mechanism for string superconductivity, which operates in models where some fermions acquire their masses from a Yukawa coupling to the Higgs field of the string [3]. Witten has shown that strings predicted in some grand unified models behave as superconducting wires. Such strings moving through magnetized cosmic plasmas can develop large currents and can give rise to a variety of astrophysical effects [4].

The particles move along the string at the speed of light. The resulting current is $J = enc$, and $dJ/dt$ is given by $dJ/dt = (ce^2/\hbar)E$. The current continues to grow until it reaches a critical value $J_c \sim emc^2/\hbar$, when $p_F = mc$. At this point particles at the Fermi level have sufficient energy to leave the string. Consequently, in this simplified picture, the growth of the current terminates at $J_c$ and the string starts producing particles at the rate $\dot{n} \sim eE/\hbar$. The fermion mass $m$ is model-dependent, but it does not exceed the symmetry breaking scale of the string, $\eta$. Hence, $J_c \leq J_{max} \sim e(\mu c^3/\hbar)^{1/2}$, where we have used the relation $\mu = \eta^2 c/\hbar$ [3].

The superconducting current in the strings is carried by charged particles which acquire a mass $M$ at the string-forming phase transition but remain massless inside the strings. These massless charge carriers move along the strings at the speed of light. The string current is bounded by a critical value, $J_c \sim eM$, at which the characteristic energy of the charge carriers become comparable to $M$, so that they have enough energy to jump out of the string. The mass $M$ is model-dependent but is limited by the string symmetry breaking scale $\eta$, $M \leq \eta$ [4].

## 2. Superconducting Cosmic Strings that Connected an Electrically Charged Black Hole

Considering the above background information the following highlights may be pointed out:

- It is quite clear that the no-hair theorem has not been completely proved yet since, it has not been proved for the Yang-Mills field and for the electrically charged black hole that have reached to critical current value ($J_c$).
- The growth of the current inside the superconducting cosmic string terminates at $J_c$ (critical current value) and the string starts producing particles.
- There is a limit to the amount of electric charge a black hole may have. Above a critical limit ($J > J_c = emc^2/\hbar$) the event horizon would be destroyed by the colossal force of electrostatic repulsion.
- An electrically charged black hole is just as likely to be positively charged as negatively charged.
- The state ($J \geq J_c$) of the electrically charged black hole may be considered as the final state of a gravitational collapse within an horizon. This means that the gravitational collapse would be destroyed in this state.

Considering the above highlights I cannot see any problem and difficulty on the connection between the superconducting cosmic strings and an electrically charged black hole in the state of ($J \geq J_c$). In addition I claim that a superconducting cosmic strings can provides the connection between two electrically charged black holes (one positively charged other negatively charged) in the same state.

Now, we can say that the superconducting cosmic strings can be connected to an electrically charged black hole, and these strings should be considered as hair of black holes. Thus, the no-hair theorem is not applicable for the electrically charged black holes.

## 3. CONCLUSION

As a conclusion, this paper described that the superconducting cosmic strings can be connected to an electrically charged black hole in the state of ($J \geq J_c$), and these strings should be considered as hair of the black holes. Thus, the no-hair theorem is not applicable for the electrically charged black holes.

## 4. REFERENCES

**[1]. J-P Luminet,** Black Holes (Cambridge University Press 1995)

**[2]. S. W. Hawking and R. Penrose,** The Nature of Space and Time (Princeton University Press 1996).

**[3]. A. Vilenkin and E.P.S. Shellard,** Cosmic Strings and Other Topological Defects (Cambridge University Press 1994).

**[4]. J.R.S. Nascimento, Inyong Cho and Alexander Vilenkin,** Charged Vacuum Condensate Near A Superconducting Cosmic String (August 29, 2002).